\documentclass[pre,twocolumn,showpacs,preprintnumbers,amsmath,amssymb]{revtex4}

\usepackage{mathrsfs}
\usepackage[dvipdfmx]{graphicx}
\usepackage{dcolumn}
\usepackage{bm}

\newcommand{\mc}{\mathcal}
\newcommand{\tl}{\tilde}
\newcommand{\ra}{\rangle}
\newcommand{\la}{\langle}
\newcommand{\tb}{\textbf}
\newcommand{\om}{\omega}
\newcommand{\Om}{\Omega}
\newcommand{\eps}{\epsilon}
\newcommand{\homg}{\hat{\omega}}

\newcommand{\hv}{\hat{v}}

\newcommand{\hFA}{\hat{F}_A}
\newcommand{\hFT}{\hat{F}_T}
\newcommand{\hxi}{\hat{\xi}}

\begin{document}

\title{Minimal Model of Stochastic Athermal Systems: Origin of Non-Gaussian Noise}

\author{Kiyoshi Kanazawa$^1$, Tomohiko G. Sano$^1$, Takahiro Sagawa$^2$, and Hisao Hayakawa$^1$}

\affiliation{
	$^1$Yukawa Institute for Theoretical Physics, Kyoto University, Kitashirakawa-oiwake cho, Sakyo-ku, Kyoto 606-8502, Japan\\
	$^2$Department of Basic Science, The University of Tokyo, Komaba, Meguro-ku, 153-8902, Japan}
\date{\today}

\begin{abstract}
		For a wide class of stochastic athermal systems, we derive Langevin-like equations driven by non-Gaussian noise, starting from master equations and developing a new asymptotic expansion. 
		We found an explicit condition whereby the non-Gaussian properties of the athermal noise become dominant for tracer particles associated with both thermal and athermal environments.
		Furthermore, we derive an inverse formula to infer microscopic properties of the athermal bath from the statistics of the tracer particle.
		We apply our formulation to a granular motor under viscous friction, and analytically obtain the angular velocity distribution function. 
		Our theory demonstrates that the non-Gaussian Langevin equation is the minimal model of athermal systems. 
\end{abstract}

\maketitle

{\it Introduction.}
	Recent developments in experimental technique has triggered both experimental and theoretical researches on fluctuations in various nonequilibrium systems~\cite{
	Bustamante,PBC,Liphardt,Collin,Hayashi,Toyabe,Carberry,Trepagnier,Blickle,Ciliberto1,Ciliberto2,Ciliberto3}. 
	The minimal model for thermally fluctuating systems is the Gaussian Langevin model, 
	which serves as a foundation of recent studies on stochastic thermodynamics~{\cite{Evans,Gallavotti,Crooks,Jarzynski,Seifert0,Kurchan,Hatano,Harada,Sagawa,Sekimoto1,Sekimoto2,Sekimoto3,Seifert1,Seifert2}}. 
	Furthermore, athermal fluctuations are also extensively studied in various systems recently~\cite{Ben-Isaac,Gov,Lin,Fodor,Eshuis,Talbot,Gnoli1,Gnoli2,Gnoli3,Gabelli,Zaklikiewicz,Blanter,Pekola,Addesso,Valenti}, 
	whose characteristic distinction from thermal fluctuations is a topic of wide interest. 
	In fact, non-Gaussian properties are experimentally observed in various athermal systems~\cite{Ben-Isaac,Gov,Lin,Fodor,Eshuis,Talbot,Gnoli1,Gnoli2,Gnoli3,Gabelli,Zaklikiewicz,Blanter,Pekola,Addesso,Valenti}, 
	and thermodynamic properties of such systems have been theoretically studied on the basis of non-Gaussian stochastic models~{\cite{Luzka,Baule,Morgado,Kanazawa1,Kanazawa2,Kanazawa3}}.

	A fundamental question then arises: {\it When and how does non-Gaussianity emerge from microscopic dynamics?} 
	While this problem has been well studied in the presence of anomalous fluctuations with asymptotically heavy-tailed distributions of waiting time or jump size~\cite{Klafter,Metzler,Klages}, 
	the origin of non-Gaussianity has not been fully understood with normal fluctuations as is the case for the conventional Langevin systems.
	Indeed, the conventional coarse-graining theories with normal fluctuations fail to explain non-Gaussian behaviors at leading order, 
	as they always produce the Gaussian noise from the central limit theorem (CLT)~{\cite{vanKampen,vanKampenB,Zwanzig,SekimotoZwanzig}}. 
	To clarify this point, let us review the theory of van Kampen~{\cite{vanKampen,vanKampenB}}. 
	In general nonequilibrium dynamics described by the master equation, the environmental noise is strongly correlated with the state of the system, 
	which implies that the noise is not white (or equivalently, state-dependent). 
	In the large system size limit, however, their correlation asymptotically disappears, and the noise distribution becomes Gaussian.  
	This is the origin of the universality of the white Gaussian noise, and is true even for the case of genuine nonequilibrium systems without the time-reversal symmetry. 
	Therefore, it is highly nontrivial to explain the origin of the non-Gaussian noise at leading order. 
	We can then rephrase the aforementioned question as follows: {\it When and how can non-Gaussianity emerge against the apparent universality of the CLT?}

	In this Letter, we answer the above question by developing a new asymptotic expansion of the master equation.  
	To leading order of the system size expansion, we derive a linear non-Gaussian Langevin equation for a wide class of athermal systems under three assumptions: 
	(i)~large system size, (ii)~strong thermal friction, (iii)~coexistence of the thermal and athermal noise.  
	Remarkably, non-Gaussianity still remains because of the violation of the CLT, though the athermal fluctuation is reduced to the white noise as the system size increases. 
	We also derive an inverse formula to infer the statistics of athermal bath from the probability distribution function (PDF) of the velocity of the system. 
	As a demonstration, we study a granular motor under viscous friction, and analytically derive its steady PDF. 
	Furthermore, we obtain a formula to estimate the velocity distribution of the surrounding granular gas from the rotor's PDF that is experimentally observed. 
	This implies that the non-Gaussianity of the PDF plays key roles to infer microscopic properties of the athermal bath. 
	
{\it Setup.}
	\begin{figure}
		\includegraphics[width=85mm]{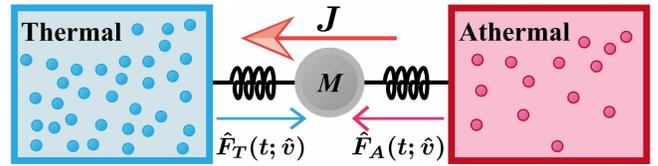}
		\caption{	(Color online) 
					Schematic of a system driven by thermal $\hat{F}_T(t;\hat v)$ and athermal $\hat{F}_A(t;\hat v)$ forces. 
					There is net energy current $J$ from the athermal to the thermal environment. 
				}
		\label{fig:schematics}
	\end{figure}
	\begin{figure*}
		\centering
		\includegraphics[width = 175mm]{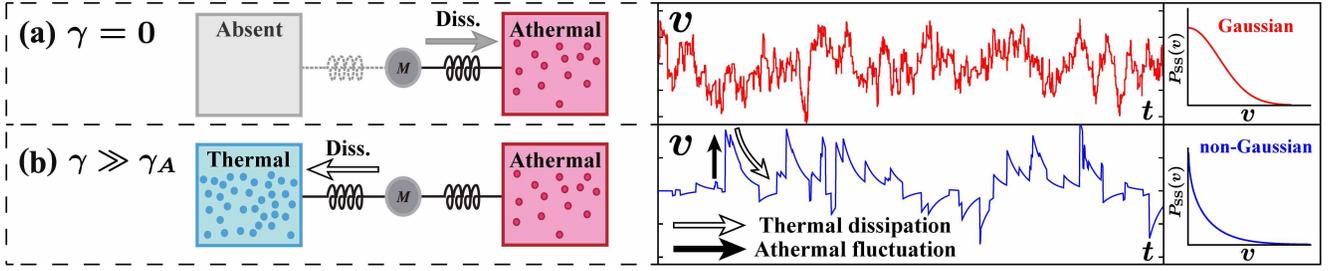}
		\caption{	(Color online)
					Typical trajectories depending on $\gamma$: 
					(a) For $\gamma=0$, the athermal force is relevant to both fluctuation and dissipation (Diss.), where the non-Gaussianity is irrelevant. 
					(b) For $\gamma\gg\gamma_A$, the athermal fluctuation is irrelevant to dissipation because of the dominance of the thermal friction. 
						The athermal non-Gaussianity then becomes relevant.
				}
		\label{fig3}
	\end{figure*}
	We consider a particle in one-dimensional space attached to thermal and athermal baths (see Fig.~{\ref{fig:schematics}} as a schematic). 
	For simplicity, we assume that the mass is unity ($M=1$) and the system obeys Markovian dynamics without mechanical potentials. 
	Then, the dynamical equation for the velocity of the particle $\hat{v}$ is written as  
	\begin{equation}
		\frac{d\hat v}{dt} =  \hat F_T(t;\hat v) + \hat F_A(t;\hat v),
	\end{equation}
	where $\hat{F}_T(t;\hat{v})$ and $\hat{F}_A(t;\hat{v})$ are stochastic forces from the thermal and athermal environments. 
	We denote the ensemble average of stochastic variable $\hat{A}$ by $\la\hat{A}\ra$.
	The thermal force $\hat{F}_T(t;\hat{v})$ is described by the sum of a linear friction and a white Gaussian noise, $\hat{F}_T(t;\hat{v})=-\gamma\hat{v}+\sqrt{2\gamma T}\hat{\xi}_G$ 
	with viscous coefficient $\gamma$, temperature $T$, and white Gaussian noise $\hat{\xi}_{G}(t)$ satisfying $\la\hat{\xi}_{G}(t_1)\ra=0$ and $\langle\hat{\xi}_{G}(t_1)\hat{\xi}_{G}(t_2)\rangle=\delta(t_1-t_2)$. 
	Here, we make a critical assumption that $\hat{F}_{A}(t;\hat v)$ is a stochastic force characterized by a small positive parameter $\epsilon$ and an $\eps$-independent Markovian jump force $\hat \eta_{A}(t;\hat v)$ such that
	\begin{equation}
		\hat F_{A}(t;\hat v) = \epsilon \hat \eta_{A}(t;\hat v), \label{eq:scaling_eps}
	\end{equation}
	where $\epsilon$ corresponds to the inverse of the system size in Refs.~{\cite{vanKampen,vanKampenB}} (see Appendix.~\ref{App1}). 
	The corresponding master equation for the velocity PDF $P(v,t)\equiv{}P(\hat{v}(t)=v)$ is given by 
	\begin{align}
		&\frac{\partial P(v,t)}{\partial t} = \gamma\left[\frac{\partial }{\partial v}v + T\frac{\partial^2 }{\partial v^2}\right]P(v,t)\notag\\
		&									+ \int_{-\infty}^\infty\!\!\!\! dy[P(v-y,t)W_{\epsilon}(v-y;y)-P(v,t)W_{\epsilon}(v;y)], \label{eq:G_ME}
	\end{align}
	where $W_{\epsilon}(v;y)$ is the athermal transition rate from $v$ with velocity jump $y$. 
	The scaling assumption~{(\ref{eq:scaling_eps})} implies that $y$ should be scaled: $\mc{Y}\equiv{}y/\epsilon$. 
	Then, the scaled transition rate $\overline{W}\left(v;\mc{Y}\right)$ for the scaled velocity jump $\mc{Y}$ satisfies 
	\begin{equation}
		\overline{W}\!\left(v;\mc{Y}\right)\!d\mc{Y}\!=\!W_{\epsilon}(v;y)dy \!\Longleftrightarrow\! W_{\epsilon}(v;y)\!=\!\frac{1}{\eps}\overline{W}\!\!\left(v;\frac{y}{\eps}\right)\!,\label{eq:scaling_assumption}
	\end{equation}
	where $\overline{W}(v;\mc{Y})$ is $\eps$-independent, corresponding to the $\eps$-independence of $\hat{\eta}_A$. 
	Note that the scaling~{(\ref{eq:scaling_assumption})} is equivalent to that introduced in Refs.~{\cite{vanKampen,vanKampenB}}, 
	and the only difference in the master equation~{(\ref{eq:G_ME})} from those in Refs.~{\cite{vanKampen,vanKampenB}} is the presence of the thermal diffusion term $L_0\equiv\gamma[(\partial/\partial{}v)v+T(\partial^2/\partial{}v^2)]$. 
	In fact, when the thermal bath is absent ($\gamma=0$) and $\hat{\eta}_A$ is stable~\cite{vanKampen} around $\hat{v}=0$, 
	the conventional Langevin equation is reproduced (see Fig.~\ref{fig3}(a)): 
	\begin{equation}
		\frac{d\hat v}{dt} = -\gamma_A\hat v + \sqrt{2\gamma_A T_A}\hxi_G, \label{eq:Gaussian_Langevin}
	\end{equation}
	where we have introduced the athermal friction $\gamma_A\equiv-\eps\alpha'_1(0)$ and temperature $2\gamma_{A}T_{A}\equiv\eps^2\alpha_2(0)$ with the Kramers-Moyal coefficient $\alpha_n(v)\equiv\int_{-\infty}^{\infty}d\mc{Y}\overline{W}(v;\mc{Y})\mc{Y}^n$. 
	We stress that the above theory is applicable to systems without microscopic reversibility, 
	which implies that microscopic irreversibility is not a sufficient condition to derive non-Gaussian models.

{\it Main results.}
	We next discuss an asymptotic expansion of Eq.~{(\ref{eq:G_ME})} in terms of the system size. 
	We make the following three assumptions, which were roughly stated in the introduction: 
	(i)~{\it Large system size: $\epsilon$ is small. }
	(ii)~{\it Strong thermal friction: $\gamma\gg\gamma_A$ (i.e., $\gamma$ is a positive constant independent of $\epsilon$). } 
	(iii)~{\it Coexistence of both thermal and athermal noise: the variance in the thermal noise is of the same order as for athermal noise (i.e., $T=\mc{T}\epsilon^2$ with an $\epsilon$-independent parameter $\mc{T}$). } 
	The condition (i) implies the weak coupling for the athermal bath, which is crucial to truncate the environmental correlation. 
	The condition (ii) implies that the thermal friction is dominant for dissipation, and the athermal force becomes irrelevant to relaxation (see Fig.~\ref{fig3}(b)). 
	We here introduce an appropriate scaled variable to remove the singularity of the small noise expansion: $\mc{V}\equiv v/\epsilon.$ 
	In the limit $\epsilon\to0$, Eq.~{(\ref{eq:G_ME})} is reduced to
	\begin{align}
		\frac{\partial \mc{P}(\mc{V},t)}{\partial t}& = \gamma\left[\frac{\partial }{\partial \mc{V}}\mc{V} + \mc{T}\frac{\partial^2 }{\partial \mc{V}^2}\right]\mc{P}(\mc{V},t)\notag\\
		&									+ \int_{-\infty}^\infty d\mc{Y}\overline{W}(0;\mc{Y})[\mc{P}(\mc{V}-\mc{Y},t)-\mc{P}(\mc{V},t)]\label{eq:G_SME}
	\end{align}
	with $\mc{P}(\mc{V},t)\equiv\epsilon{}P(v,t)$.  
	Remarkably, $\overline{W}(0;\mc{Y})$ is independent of the system's velocity $\hat{\mc{V}}$, 
	which implies that the environmental correlation disappears and the athermal fluctuation is reduced to the white noise. 
	Furthermore, the non-Gaussianity still remains after this reduction, as seen from the system's steady distribution (see Fig.~\ref{fig3}(b)). 	
	This is the violation of the CLT.  
	Then, Eq.~{(\ref{eq:G_SME})} is equivalent to the Langevin-like equation with a white non-Gaussian noise term~{\cite{vanKampenB}}: 
	\begin{equation}
		\frac{d\hat{\mc{V}}}{dt} = -\gamma\hat{\mc{V}} + \sqrt{2\gamma\mc{T}}\hat \xi_{G} + \hat \xi_{NG},\label{eq:NGL}
	\end{equation}
	where $\hat{\xi}_{NG}$ is the white non-Gaussian noise with transition rate $\overline{W}(0;\mc{Y})$. 
	This is the first main result of this Letter. 
	We stress that Eq.~{(\ref{eq:NGL})} is exactly solvable~\cite{Eliazar}. 
	Indeed, Eq.~{(\ref{eq:G_SME})} is reduced to $(d/ds)\tl{P}_{\rm SS}(s)=(\Phi(s)/\gamma{}s)\tl{P}_{\rm SS}(s)$ in the steady state, 
	where the convolution in Eq.~{(\ref{eq:G_SME})} is simplified 
	by introducing the steady PDF $\mc{P}_{\rm SS}(\mc{V})\equiv\lim_{t\to\infty}\mc{P}(\mc{V},t)$, its Fourier representation $\tl{P}_{\rm SS}(s)\equiv\int_{-\infty}^{\infty}d\mc{V}e^{is\mc{V}}\mc{P}_{\rm SS}(\mc{V})$, 
	and the cumulant function $\Phi(s)\equiv\int_{-\infty}^{\infty}d\mc{Y}\overline{W}(0;\mc{Y})(e^{is\mc{Y}}-1)-\gamma\mc{T}s^2$. 
	This equation is easily solved as $\tl{P}_{\rm SS}(s)=\exp[\int_0^sds'\Phi(s')/\gamma{}s']$. 
	The stationary PDF is then given by $\mc{P}_{\rm SS}(\mc{V}) = \int_{-\infty}^{\infty}ds\exp{\left[-is\mc{V}+\int_{0}^s ds'\Phi(s')/\gamma s'\right]}/2\pi.$ 
	We further obtain the inverse formula of the transition rate $\overline{W}(v;\mc{Y})$ from the stationary PDF as
	\begin{equation}
		\overline{W}(0;\mc{Y}) \!=\! \gamma\!\int_{-\infty}^\infty \!\frac{ds}{2\pi}e^{-is\mc{Y}}\!\!\left[\!\lambda^* \!+\! \mc{T}s^2\!+\!s\frac{d}{ds}\log{\tl{P}_{\rm SS}}(s)\!\right]\!\!,\label{eq:Exp_Dst}
	\end{equation}
	where we have introduced $\lambda^*\equiv\int_{-\infty}^{\infty}d\mc{Y}\overline{W}(0;\mc{Y})/\gamma$. 
	Note that $\lambda^*=-\lim_{s\to\infty}[s(d/ds)\log{\tl{P}_{SS}(s)}+\mc{T}s^2]$ according to the Riemann-Lebesgue lemma~\cite{RLlemma}. 
	This is the second main result of this Letter, which connects the microscopic transition rate $\overline{W}(v;\mc{Y})$ and the observable $\mc{P}_{\rm SS}(\mc{V})$.	
	Equation~{(\ref{eq:Exp_Dst})} is derived from the inverse Fourier transformation of the definition of the cumulant function as 
	$\overline{W}(0;\mc{Y})=\gamma\int_{-\infty}^{\infty}dse^{-is\mc{Y}}(\lambda^*+\mc{T}s^2+\Phi(s)/\gamma)/2\pi$ and the relation $\Phi(s)/\gamma=s(d/ds)\log{\tl{P}_{\rm SS}(s)}$.
	The effectiveness of Eq.~{(\ref{eq:Exp_Dst})} will be demonstrated later by an example of a granular motor.
	We note that our formulation is applicable to the small noise expansion for a single multiplicative L\'evy noise~{\cite{Albeverio}}. 
	We also note that our formulation reduces to the independent kick model~{\cite{Talbot,Gnoli1,Gnoli2,Gnoli3}} in the limit $\gamma\to\infty$ (see Appendix.~\ref{App2}).
	
	Equation~{(\ref{eq:NGL})} does not satisfy the detailed balance condition, 
	because there is net energy current from the athermal to the thermal environment as 
	$J=\la{}d\hat Q/dt\ra_{\rm SS}=K_2>0,$
	where we have introduced the second cumulant $K_2\equiv(d^2/d(is)^2)\Phi(s)|_{s=0}$ and the heat current~{\cite{Sekimoto1,Sekimoto2,Sekimoto3,Kanazawa1}} $d\hat{Q}/dt=(\gamma\hat{\mc{V}}-\sqrt{2\gamma\mc{T}}\hat\xi_G)\circ\hat{\mc{V}}$
	with the product defined in the Stratonovich sense~{\cite{Gardiner}}. 
	Remarkably, the direction of heat current is independent of the thermal temperature $\mc{T}$, 
	which implies that the athermal environment has high effective temperature under the assumptions (i-iii). 
	This result is consistent with various experiments~\cite{Ben-Isaac,Gnoli1,Gnoli2,Gnoli3,Gabelli,Zaklikiewicz}, where effective temperatures of athermal noise are reported to be much higher than the room temperature.
		
	We now discuss the physical criteria behind assumption~{(iii)}. 
	Let us first expand $W_{\epsilon}(v;y)$ as $W_{\epsilon}(v;y^*)=W_{\epsilon}(0;y^*)+W_{\epsilon}^{(1)}(0;y^*)v+O(v^2)$ with $W_{\epsilon}^{(n)}(v;y^*)\equiv{}\partial^nW_{\epsilon}(v;y^*)/\partial{}v^n$ with the typical velocity jump $y^*$.
	The essence of our expansion is that the non-linear part of $W_{\epsilon}(v;y)$ is asymptotically irrelevant in the small $\epsilon$ limit as shown in Eq.~{(\ref{eq:G_SME})}, i.e. $|W_{\epsilon}(0;y)| \gg v |W_{\epsilon}^{(1)}(0;y)|$. 
	We then introduce the non-linear temperature $T_{NL}\equiv|W_{\epsilon}(0;y^*)/W_{\epsilon}^{(1)}(0;y^*)|^2/2$, 
	which characterizes the relevance of the non-linear part of $W_{\epsilon}(v;y)$.
	Then, the assumption~(iii) is equivalent to $T/T_{NL}=O(\epsilon^2)\ll{}1$. 
 	
 {\it Violation of the CLT.}
	We here discuss the mechanism of the violation of the CLT. 
	According to the CLT, the summation of the independent and identically-distributed (i.i.d.) variables converges to a Gaussian random variable. 
	Since the white noise $\hxi$ is regarded as i.i.d., the summation $\int_0^t ds\hxi(s)/\sqrt{t}$ converges to a Gaussian variable for $t\gg \tau_P$, 
	where $\tau_P$ is the characteristic time interval between athermal collisions. 
	When the thermal friction is absent, the relaxation time $\tau_R$ diverges because the athermal friction is proportional to $\eps$ as $\gamma_A=-\eps\alpha_1'(0)$, 
	which ensures that the system moves slowly in the time scale of $\tau_P$ and the CLT is applicable. 
	In contrast, when the thermal friction is sufficiently strong, $\tau_R$ is the same order of $\tau_P$ ($\tau_R\simeq\tau_P$). The CLT is not applicable anymore in this situation. 
	The above picture clarifies the mechanisms of the violation of the CLT and the emergence of non-Gaussianity. 

\begin{figure}
		\centering
		\includegraphics[width=83mm]{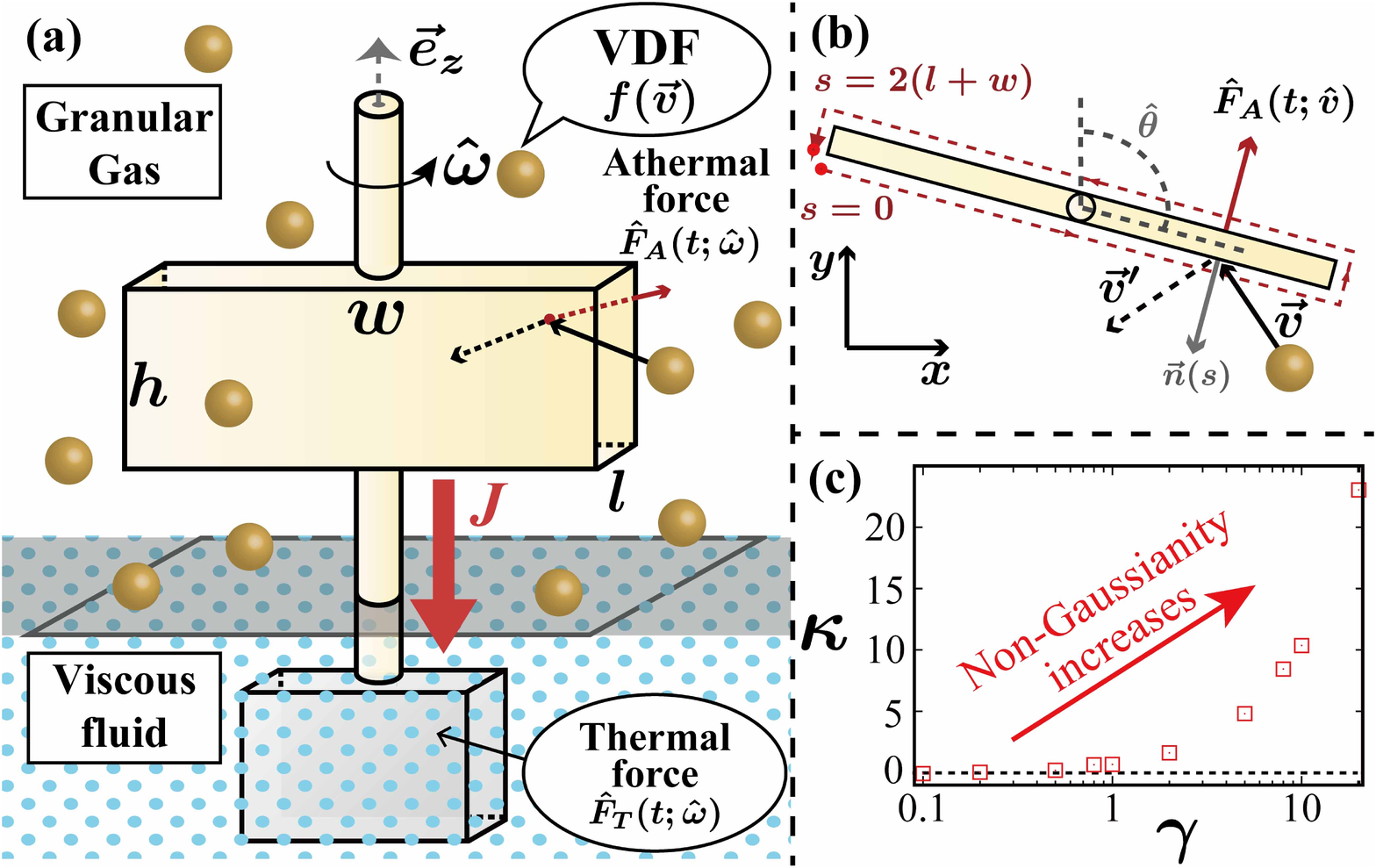}
		\caption{	(Color online)
					(a) Schematic of a rotor driven by the thermal force $\hFT(t;\homg)$ from the viscous fluid and the athermal force $\hFA(t;\homg)$ caused by collisions of the granular particles. 
						The heat current $J$ flows from the granular gas to the viscous fluid. 
						We fix parameters as $M=I=h=e=\rho=v_0=1$, $T=l=0$, $m=0.01$, and $w=\sqrt{12}$ for numerical simulations.
					(b) Collisional rules are illustrated. 
					(c) Numerical demonstration of the emergence of $\kappa\equiv\la\homg^4\ra/\la\homg^2\ra^2-3$ corresponding to the increases of the viscosity $\gamma$.
				}
		\label{fig:gmotor}
	\end{figure}

{\it Example: granular motor under viscous friction.} 
	Let us consider a granular motor under viscous friction (see Fig.~{\ref{fig:gmotor}(a)}). 
	We prepare a rotor of cuboid shape with mass $M$, inertial moment $I$, height $h$, width $w$, and depth $l$. 
	The rotor is immersed in two environments: a viscous fluid and a granular gas. 
	The viscous fluid is a thermal bath characterized by viscous coefficient $\gamma$ and temperature $T$. 
	The granular gas under vertical vibration is an steady athermal bath characterized by velocity distribution function (VDF) $f(\vec v)$, particle's mass $m$, and restitution coefficient $e$. 
	For simplicity, we assume an exponential granular VDF~\cite{Olafsen} as $f(\vec{v})=e^{-|\vec{v}|/v_0}/8\pi v_0^3$ and $T=l=0$. 
	Note that similar setups under dry friction are experimentally realized in Refs.~{\cite{Eshuis,Gnoli1,Gnoli2,Gnoli3}}. 
	The angular velocity $\hat{\omega}$ is driven by the thermal and athermal forces in the viscous fluid and the granular gas, respectively. 
	We assume that the granular gas is so dilute that the athermal force $\hFA$ can be described by the Boltzmann-Lorentz model~{\cite{Talbot,Eshuis,Gnoli1,Gnoli2,Gnoli3,Lorentz1905,Gallavotti1969,Sphon1978,Brey,Brilliantov}}. 
	By introducing the mass ratio $\epsilon\equiv{}m/M$, we obtain the master equation for $\hat{\omega}$ as 
	\begin{align}
		&\frac{\partial }{\partial t}P(\omega,t) = \gamma \left[\frac{\partial }{\partial \omega}\omega +\frac{T}{I}\frac{\partial^2 }{\partial \omega^2}\right]P(\omega,t)\notag\\
		&+\!\int \!dy\!\left[P(\omega\!-\!y,t)W_\eps(\omega\!-\!y;y)\!-\!P(\omega,t)W_\eps(\omega;y)\right],\label{BLeq}
	\end{align}
	where we have introduced the angular velocity PDF $P(\omega,t)\equiv{}P(\hat \omega(t)=\omega)$,
	the athermal transition rate $W_\eps(\omega; y)=\rho{}h\int{}ds\int d\vec{v}f(\vec{v})\Theta(\Delta\vec{V}\cdot\vec{n})|\Delta\vec{V}\cdot\vec{n}|\delta[y-\Delta\omega]$, 
	the coordinate along the cuboid $s$, the normal unit vector to the surface $\vec n(s)$, the number density $\rho$, the inertia radius $R_I\equiv \sqrt{I/M}$, $\vec{e}_z\equiv(0,0,1)$,
	$\vec{V}(s)\equiv\omega\vec{e}_z\times\vec{r}(s)$, 
	$\Delta\vec{V}(s)\equiv\vec{V}(s)-\vec v$, $\vec{t}(s)\equiv\vec{e}_z\times\vec{n}(s)$, $g(s)\equiv \vec r(s)\cdot \vec t(s)/R_I$, 
	and $\Delta\omega\equiv\eps{}g(s)(1+e)(\Delta\vec{V}(s)\cdot\vec{n})/R_I(1+\epsilon{}g^2(s))$ (see Fig.~{\ref{fig:gmotor}(b)}).  
	We stress that the granular force is not white noise in general because of the presence of the environmental correlation. 
	Indeed, the athermal transition rate $W_\eps(\om;y)$ depends on $\omega$. 
	We also stress that the non-Gaussianity $\kappa\equiv\la\homg^4\ra/\la\homg^2\ra^2-3$ is irrelevant for $\eps\to 0$ 
	when the thermal friction is absent ($\gamma=0$) as shown in Fig.~\ref{fig:gmotor}(c),
	though Eq.~{(\ref{BLeq})} has no time-reversal symmetry~\cite{Brey,Brilliantov}. 
	
\begin{figure}
		\includegraphics[width=85mm]{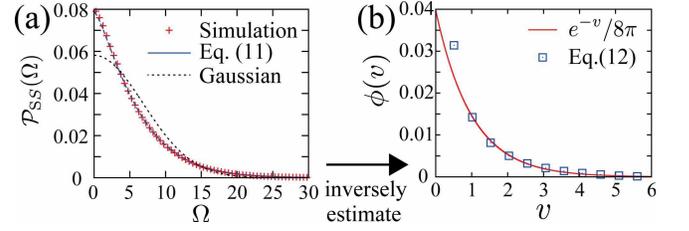}
		\caption{	(Color online)
					(a)~Steady PDF of the angular velocity $\hat\omega$ obtained from MCS of the Boltzmann-Lorentz equation~{(\ref{BLeq})} for $\gamma=2$ (the cross points),
						our theoretical line~{(\ref{dist_Gaussian})} (the solid line), and the conventional Gaussian theory~{(\ref{eq:Gaussian_Langevin})} (the dashed line).  
					(b)~The granular VDF estimated from $\mc{P}_{\rm{SS}}(\Omega)$ using Eq.~{(\ref{eq:infer_granular})}. 
						Note that the accuracy of the point at $v\simeq 0.5$ is not good because of the singularity in Eq.~{(\ref{eq:infer_granular})} around $v=0$. 
				}
		\label{fig:infer}
	\end{figure}

	Here we assume that the mass ratio $\epsilon$ is small and the thermal friction is much larger than the athermal friction (i.e., $\eps$-independence of $\gamma$). 
	By introducing a scaled variable $\hat{\Omega}\equiv\hat{\omega}/\epsilon$, 
	we obtain the non-Gaussian Langevin equation in the limit $\epsilon\rightarrow+0$:
	\begin{equation}
		\frac{d\hat{\Omega}}{dt} = -\gamma \hat \Omega + \hat \eta_g,\label{eq:Lang_grnmtr}
	\end{equation}
	where $\hat{\eta}_g$ is the granular collisional torque characterized by 
	the cumulant function $\Phi(s)\equiv-\rho{}hwv_0\Om_g^2s^2(5+3\Om_g^2s^2)/2(1+\Om_g^2s^2)^2$ with $\Om_g\equiv{}wv_0(1+e)/2R_I^2$. 
	We then obtain the exact steady PDF for the scaled angular velocity $\tl{\Omega}\equiv\Omega/\Omega_g$ (see Appendix.~\ref{App3}):
	\begin{equation}
		\mc{P}_{\rm SS}(\tl{\Omega}) 
		=\int_{-\infty}^\infty \frac{ds}{2\pi}\frac{e^{[-is\tl{\Omega} - v_0 s^2/\tl{v}(1+s^2)]}}{(1+s^2)^{3v_0/2\tl{v}}},\label{dist_Gaussian} 
	\end{equation}
	where $\tl{v}\equiv2\gamma/\rho{}hw$. 
	The validity of Eq.~{(\ref{dist_Gaussian})} is numerically checked by the Monte Carlo simulation (MCS) of Eq.~{(\ref{BLeq})} shown in Fig.~{\ref{fig:infer}}(a), 
	where the theoretical line perfectly agrees with the numerical data, while the conventional Gaussian model~{(\ref{eq:Gaussian_Langevin})} does not. 
	Note that the granular impulses are reduced to the white noise as the environmental correlation disappears (i.e., the athermal force $\hat{\eta}_g$ becomes $\hat{\om}$-independent). 
	Furthermore, the non-Gaussianity becomes relevant as the thermal friction increases as illustrated in Fig.~\ref{fig:gmotor}(c). 
	We also note that the steady heat current $J=I\gamma\la\hat{\Omega}^2\ra>0$ flows from the granular gas to the viscous fluid, 
	which implies that the rotor is far from thermal equilibrium.
	
	We demonstrate the usefulness of the inverse formula~{(\ref{eq:Exp_Dst})} to infer the properties of non-equilibrium baths. 
	We assume that the VDF of the granular gas is isotropic: $f(\vec v)=\phi(|\vec v|)$.  
	From Eq.~{(\ref{eq:Exp_Dst})}, we obtain the following formula for an arbitrary $\phi(v)$:	 
	\begin{equation}
		\phi(v) =  \int_{0}^\infty \!\!\!\frac{ds}{\pi |v|}\left[a-\frac{bs^2}{2}-cs^3\frac{d}{ds}\log{\tl{P}_{SS}(s/F_g)}\right]\cos{(sv)},\label{eq:infer_granular}
	\end{equation}
	where $a\equiv\int_{-\infty}^{\infty}dv|v|\phi(v)$, $b\equiv\int_{-\infty}^{\infty}dv|v|^3\phi(v)$, $c\equiv\gamma/2\pi\rho{}hw$, $F_g\equiv{}w(1+e)/2R_I^2$, and $\tl{P}_{\rm SS}(s)\equiv{}\int_{-\infty}^{\infty}d\Om{}e^{is\Om}\mc{P}_{\rm SS}(\Om)$. 
	Equation~{(\ref{eq:infer_granular})} is applicable to infer the granular VDF from the observation of the rotor's PDF. 
	This implies that the non-Gaussianity in $\mc{P}_{\rm SS}(\Omega)$ is useful to infer the microscopic properties of the athermal bath. 
	Note that the coefficients $a$ and $b$ are determined by the Riemann-Lebesgue lemma~{\cite{RLlemma}} $\lim_{s\rightarrow\infty}[a-bs^2/2-cs^3(d/ds)\log{\tl{P}}_{SS}(s/F_g)]=0$.
	We demonstrate the validity of Eq.~{(\ref{eq:infer_granular})} for $\phi(v)=e^{-|v|}/8\pi$ in Fig.~{\ref{fig:infer}(b)}, 
	where $\phi(v)$ is estimated from Eq.~{(\ref{eq:infer_granular})}. 
	This is a clear demonstration of the effectiveness of the inverse formula~{(\ref{eq:infer_granular})}. 
	
	The meaning of the inverse formula~{(\ref{eq:infer_granular})} can be understood from the viewpoint of ``cooling" of the rotor. 
	In the absence of a thermal environment, the rotor's effective temperature approaches that of the granular gas. 
	Conversely, in the presence of a thermal environment, the effective temperature is less than that of the granular gas, because the thermal environment plays the role of a ``cooler." 
	It absorbs redundant information from the rotor's motion, and therefore, the precise information of athermal noise (i.e., high-order cumulants) is accessible from the rotor's dynamics. 
	
{\it Conclusion.}
	In this Letter, we have considered a tracer particle attached to both thermal and athermal environments, and derived a non-Gaussian Langevin equation~{(\ref{eq:NGL})} 
	subject to the condition that the athermal stochastic force is irrelevant during relaxation. 
	We also derived an inverse formula~{(\ref{eq:Exp_Dst})} to infer the environmental information from the observation of the tracer particle. 
	We applied our formulation to a granular motor under viscous friction, 
	and analytically obtained the stationary PDF~{(\ref{dist_Gaussian})} and an inverse formula~{(\ref{eq:infer_granular})} on the granular velocity distribution. 

	We revealed the emergence of the non-Gaussianity and its microscopic origins. 
	Extensions of this formulation to multidimensional non-linear systems are planned for the future. 
	Our theory serves as a foundation of athermal statistical mechanics and would be important for various fields of science, such as biophysics, chemistry, and econophysics.

	We are grateful for useful discussion with N. Nakagawa, K. Sekimoto, and A. Puglisi. 
	This work was supported by the JSPS Core-to-Core Program ``Non-equilibrium dynamics of soft matter and information,"
	Grants-in-Aid for the Japan Society for Promotion of Science (JSPS) Fellows (Grant Nos. 24$\cdot$3751 and 26$\cdot$2906), 
	and JSPS KAKENHI Grant Nos. 25287098 and 25800217.

\appendix

\section{Review of van Kampen's microscopic theory}\label{App1}
	We here review the microscopic derivation of the Gaussian Langevin equation. 
	Let us consider a system driven by a single stochastic environment as 
	\begin{equation}
		\frac{d\hv}{dt} = \hFA,
	\end{equation}
	where $\hFA$ is a stochastic force from the single environment. 
	We here make a critical assumption that $\hFA$ is scaled by a positive small number $\eps$ as 
	\begin{equation}
		\hFA = \eps \hat{\eta}_A(t;\hv), \label{spp:assmp:scale_SDE}
	\end{equation}
	where $\hat{\eta}_A$ is a Markovian jump force independent of $\eps$. 
	Let us write the jump rate of $\hat{\eta}_A(t;\hv)$ as $\overline{W}(v;\mc{Y})$, 
	where $\overline{W}(v;\mc{Y})$ is the $\eps$-independent transition probability per unit time on the condition $\hat v(t)=v$ with the amplitude of the Poisson noise $\mc{Y}$. 
	We assume that the master equation for the velocity $\hat v$ is given by
	\begin{equation}
		\frac{\partial P(v,t)}{\partial t} = \int_{-\infty}^\infty dy[P(v-y,t)W_\eps(v-y;y)-P(v,t)W_\eps(v,y)],\label{spp:eq:master_eq}
	\end{equation}
	where $P(v,t)\equiv P(\hv(t)=v)$ is the probability distribution function (PDF) for the system's velocity and $W_\eps(v;y)$ is the transition rate for $\hv$ on the condition $\hv(t)=v$ with velocity jump $y$. 
	Considering the relation~{(\ref{spp:assmp:scale_SDE})}, $y$ and $\mc{Y}$ are connected by $y=\eps \mc{Y}$. 
	Then, the Jacobian relation holds as 
	\begin{equation}
		W_\eps(v;y)dy = \overline{W}(v;\mc{Y})d\mc{Y} \Longleftrightarrow W_\eps(v;y)=\frac{1}{\eps}\overline{W}\left(v;\frac{y}{\eps}\right). \label{spp:eq:scaling_kernel}
	\end{equation}
	The Kramers-Moyal expansion for the master equation is given by
	\begin{equation}
		\frac{\partial P(v,t)}{\partial t} = \sum_{n=1}^\infty \frac{(-\eps)^n}{n!}\frac{\partial^n}{\partial v^n}[\alpha_n(v)P(v,t)],\label{spp:eq:Kramers_Moyal}
	\end{equation}
	where we have introduced the Kramers-Moyal coefficient
	\begin{equation}
		\alpha_n(v) \equiv \int_{-\infty}^\infty d\mc{Y}\overline{W}(v;\mc{Y})\mc{Y}^n. 
	\end{equation}
	We here assume the stability condition for the noise around $\hv=0$: 
	\begin{equation}
		\alpha_1^{(0)} = 0, \>\>\> \alpha^{(1)}_1 = -\bar{\gamma}_A < 0, \>\>\> \alpha_2^{(0)}=2\bar{\gamma}_A\bar{T}_A > 0,
	\end{equation}
	where $\alpha_1(v)$ has a single zero point and we introduce the expansion $\alpha_n(v)=\sum_{k=0}^\infty v^k\alpha_n^{(k)}/k!$. 
	By introducing the following scaled variables as $\tau = \eps t$, $V = v/\sqrt{\eps}$, Eq.~{(\ref{spp:eq:Kramers_Moyal})} is reduced to
	\begin{align}
		&\frac{\partial P(V,\tau)}{\partial \tau} 
												= \bigg[\frac{\partial }{\partial V}\bar{\gamma}_AV - \sum_{k=2}^\infty \frac{\eps^{(k-1)/2}}{k!}\frac{\partial }{\partial V}\alpha_1^{(k)}V^k\notag\\
												&+\sum_{n=2}^\infty\sum_{k=0}^\infty \frac{(-1)^n\eps^{(n+k)/2-1}}{n!k!}\frac{\partial^n}{\partial V^n}V^k\alpha_n^{(k)}\bigg]P(V,\tau)\notag\\
												 &= \bar{\gamma}_A\left[\frac{\partial }{\partial V}V + \bar{T}_A\frac{\partial^2 }{\partial V^2}\right]P(V,\tau) + O(\eps^{1/2}),\label{spp:eq:derived_FP}
	\end{align}
	which is equivalent to the Gaussian Langevin equation~{(5)} in the main text.
	
	We here note that the scaling~{(\ref{spp:eq:scaling_kernel})} is essentially equivalent to that introduced by van Kampen~\cite{vanKampen,vanKampenB,Gardiner}, 
	where $\eps$ corresponds to the inverse of the system size $\Omega_{\rm sys}$ as $\Omega_{\rm sys}\equiv 1/\eps$. 
	To see this point, let us transform velocity variables
	\begin{equation}
		a' \equiv \frac{v}{\eps}=\Omega_{\rm sys} v, \>\>\> \Delta a \equiv \frac{y}{\eps} = \Omega_{\rm sys} y.
	\end{equation}
	Then, the scaling assumption~{(\ref{spp:eq:scaling_kernel})} is equivalent to 
	\begin{equation}
		W(a';\Delta a) = \Omega_{\rm sys} \overline{W}\left(\frac{a'}{\Omega_{\rm sys}}, \Delta a\right),\label{spp:eq:scaling_kernel_Omega}
	\end{equation}
	where we have introduced the transition rate for the transformed variable $a'$ as $W(a';\Delta a)\equiv W_\eps(v;y)$. 
	The scaling~{(\ref{spp:eq:scaling_kernel_Omega})} is exactly equal to that introduced in Ref.~\cite{Gardiner} on page 277. 
	
	\subsection*{Note on the invalidity of the Langevin description for rare trajectories}
	We here note that the Langevin equation is an effective description only for typical trajectories, where one does not observe rare trajectories. 
	In the above discussion, we have implicitly assumed that the scaled velocity $V$ is not much larger than the typical velocity $V^*\equiv \sqrt{\overline{T}_A}$ (i.e., $|V|\lesssim V^*$). 
	In fact, when $V$ is much larger than $V^*$ as $V/V^* = O(1/\eps)$, all of the terms on the right hand side of Eq.~{(\ref{spp:eq:derived_FP})} are relevant, which implies the invalidity of the system size expansion for $|V|\gg V^*$ 
	(i.e., the distribution of the tail cannot be well described by the Langevin description in general). 
	This is because the system size expansion is not a uniform asymptotic expansion in terms of the velocity. 
	Fortunately, however, the probability of such rare trajectories is estimated to be extremely small and is irrelevant to averages of ordinary physical quantities. 
	The Langevin description is proved to be effectively valid in this sense. 
	
	\subsection*{Note on the validity of the van Kampen's theory for systems without microscopic reversibility}
	We note that the system size expansion is valid for genuine nonequilibrium systems without time-reversal symmetry because it is based on the central limit theorem. 
	Indeed, the effectiveness of the Gaussian Langevin description is discussed for some granular systems in Refs.~\cite{Brey,Brilliantov}, 
	where the starting points of the models are given by master equations without time-reversal symmetry. 
	We also note that these results do not contradict the time-reversal symmetry of the Langevin model, because the Langevin model is just an effective description for typical trajectories. 
	Even if the system is well-described by the Langevin equation for typical trajectories, the time-irreversal symmetry can be observed in general for rare trajectories. 
	
	\subsection*{Note on the state-dependence of noise}
	We here note that the fluctuation described by the master equation~{(\ref{spp:eq:master_eq})} is state-dependent noise, which is not simple white noise. 
	In fact, the transition rate $W_\eps(v;y)$ for the velocity jump $y$ depends on $v$, 
	which implies the strong correlation between the system and the environment. 
	We also note that the state-dependent noise cannot be generally written as a single multiplicative noise, 
	because the time-series of the Poisson flights is independent of the state of the system for the single multiplicative noise.

\section{Relation to the independent kick model}\label{App2}
	The non-Gaussian Langevin equation reproduces the independent kick model in the strong friction limit. 
	The independent kick was originally introduced to explain the behavior of the granular motor in the presence of the dry friction~{\cite{Talbot,Gnoli1,Gnoli2,Gnoli3}}. 
	According to the main text, the steady distribution of the non-Gaussian Langevin equation is represented in the Fourier space as  
	\begin{equation}
		\tl{P}_{\rm SS}(s) = \exp{\left[\int_0^s \frac{ds'\Phi(s')}{\gamma s'}\right]}, \label{spp:g_Fdist}
	\end{equation}
	where the cumulant function is given by
	\begin{equation}
		\Phi(s)=-\gamma \mc{T}s^2+\int_{-\infty}^\infty d\mc{Y}\overline{W}(0;\mc{Y})(e^{i\mc{Y}s}-1)
	\end{equation}
	with the Poisson transition rate $\overline{W}(0;\mc{Y})$.
	Let us consider the case without the Gaussian part ($\mc{T}=0$). 
	In the strong friction limit $\gamma \rightarrow \infty$, Eq.~{(\ref{spp:g_Fdist})} is reduced to
	\begin{align}
		\tl{P}_{\rm SS}(s) 	&= 1 + \int_0^s \frac{ds'\Phi(s')}{\gamma s'} + O(\gamma^{-2})\notag\\
											&= 1 + \int_{-\infty}^\infty d\mc{Y}\int_0^s ds' \frac{\mc{W}(\mc{Y})}{\gamma s'}(e^{is'\mc{Y}}-1) + O(\gamma^{-2}).\label{spp:eq:independent}
	\end{align}

	On the other hand, the system is kicked by rare collisions and instantly relaxes to the rest state in the independent kick model~{\cite{Talbot,Gnoli1,Gnoli2,Gnoli3}}. 
	This implies the following scenario. The system is typically in the rest state $\mc{V}=0$. 
	However, an occasional collision at time $t=0$ with a Poisson flight distance $\mc{Y}$ changes the state from $\mc{V}(-0)=0$ to $\mc{V}(+0)=\mc{Y}$, 
	and the system freely relaxes as $\mc{V}(t)=\mc{Y}e^{-\gamma t}$. 
	This scenario leads to 
	\begin{equation}
		\la h(\hat{\mc{V}})\ra \simeq \int_{-\infty}^\infty d\mc{Y}\overline{W}(0;\mc{Y})\int_0^\infty dth(\mc{V}(t)),\label{spp:eq:IKM_general}
	\end{equation}
	where $h(v)$ is an arbitrary function.
	Substituting $h(\mc{V})=e^{is\mc{V}}-1$ into Eq.~{(\ref{spp:eq:IKM_general})}, we obtain 
	\begin{align}
		\tl{P}_{\rm SS}(s) - 1 	&\simeq \int_{-\infty}^\infty d\mc{Y}\overline{W}(0;\mc{Y})\int_0^\infty dt\left\{\exp{\left[is\mc{Y}e^{-\gamma t}\right]}-1\right\}\notag\\
								&= \int_{-\infty}^\infty d\mc{Y}\overline{W}(0;\mc{Y})\int_0^\infty dt\sum_{n=1}^\infty \frac{1}{n!}(is\mc{Y})^ne^{-\gamma nt}\notag\\
								&= \int_{-\infty}^\infty d\mc{Y}\overline{W}(0;\mc{Y})\sum_{n=1}^\infty \frac{(is\mc{Y})^n}{n\gamma n!}\notag\\
								&= \int_{-\infty}^\infty d\mc{Y}\int_0^s ds'\frac{\overline{W}(0;\mc{Y})}{\gamma s'}(e^{is'\mc{Y}}-1),
	\end{align}
	which is equivalent to Eq.~{(\ref{spp:eq:independent})}. 
	Thus, our theory is equivalent to the independent kick model in the strong friction limit.

\section{Granular motor under the viscous and dry frictions}\label{App3}
	\subsection{Setup}
		We consider a granular motor under the viscous friction. 
		The motor is a cuboid of height $h$, width $w$, and length $l$ in the granular gas as in Fig~{\ref{fig:gmotor}(a)}. 
		The cuboid rotates around the $z$-axis, and the rotational angle $\hat \theta$ fluctuates because of collisional impacts by surrounding granular particles. 
		We assume that there exists the Coulombic friction during the rotation around the axis. 
		Let us first consider its collision rules (see Fig.~{\ref{fig:gmotor}(b)}). 
		We assume that the motor and a particle collide at the position $\vec r$.
		We denote the motor's angular velocity and particle's velocity by $\omega$ and $\vec v$, respectively. 
		The moment of inertia along the $z$-axis and the radius of inertia are respectively given by $I$ and $R_I\equiv \sqrt{I/M}$. 
		The conservation of the angular momentum and the definition of the restitution coefficient $e$ are respectively given by 
		\begin{equation}
			I\omega \vec e_z + m \vec r\times \vec v = I\omega' \vec e_z + m \vec r\times \vec v',\>\>\>\>
			-\frac{(\vec V' -\vec v')\cdot \vec n}{(\vec V-\vec v)\cdot \vec n} = e,\label{am_dr}
		\end{equation}
		where $\omega'$, $\vec V'$ and $\vec v'$ are the angular velocity of the motor, the velocity of the motor and the velocity of the particle after the collision, respectively,
		and $\vec n$ is the normal unit vector on the surface, and $\vec e_z \equiv (0,0,1)$.
		We assume the non-slip condition for the collision: the velocity change of the particle is perpendicular to the surface as
		\begin{equation}
			\vec v' = \vec v +\beta \vec n\label{nsc}
		\end{equation}
		with an appropriate coefficient $\beta$. 
		We note the following relations: 
		\begin{equation}
			\vec V = \omega \vec e_z \times \vec r,\>\>\> \vec V' = \omega' \vec e_z \times \vec r.\label{r_v_av}
		\end{equation}
		Solving Eqs.~{(\ref{am_dr})}, {(\ref{nsc})}, and~{(\ref{r_v_av})}, we obtain 
		\begin{equation}
			\Delta \omega\equiv  (1+e)\frac{\Delta \vec V\cdot \vec n}{R_I}\frac{\epsilon (\vec r\cdot \vec t/R_I)}{1+\epsilon (\vec r\cdot \vec t/R_I)^2},\>\>\>
			\beta = \frac{(1+e)\Delta \vec V\cdot \vec n}{1+\epsilon (\vec r\cdot \vec t/R_I)^2},\label{col_rule}
		\end{equation}
		where we introduced $\vec t\equiv \vec e_z \times \vec e$, $\Delta \vec V\equiv \vec V-\vec v$, and $\epsilon\equiv m/M$. 
		Based on the collision rule~{(\ref{col_rule})}, we model this setup as the Boltzmann-Lorentz equation~{\cite{Talbot,Gnoli1,Gnoli2,Gnoli3,Brilliantov}}: 
		\begin{align}
			&\frac{\partial }{\partial t}P(\omega,t) = \gamma\left[\frac{\partial }{\partial \omega} \omega +\frac{T}{I}\frac{\partial^2 }{\partial \omega^2}\right]P(\omega,t) \notag\\
			&	+\int \!\!dy\left[P(\omega\!-\!y,t)W_\eps(\omega\!-\!y;y)\!-\!P(\omega,t)W_\eps(\omega;y)\right],\label{eq:Suppl:BLeq}\\
			&W_\eps(\omega ;y) \!=\!\rho h\!\int\! ds\! \int\! d\vec v f(\vec v)\Theta(\Delta \vec V\cdot \vec n)|\Delta \vec V\cdot\vec n|\delta(y-\Delta \omega),
		\end{align}
		where $s$ is the coordinate along the cuboid, $f(\vec v)$ is the granular distribution function, $\gamma$ is the coefficient of the viscous friction, 
		$\vec n(s)$ is the normal unit vector to the surface at $s$, and we have introduced 
		\begin{align}
			&\vec V(s) \equiv \omega\vec e_z \times \vec r(s), \>\> g(s)\equiv \frac{\vec r(s)\cdot \vec t(s)}{R_I}, 
			\>\> \vec t(s)\equiv \vec e_z\times \vec n(s), \notag\\
			&\Delta \vec V(s) \equiv \vec V(s)-\vec v,
			\>\> \Delta \omega \equiv \frac{\Delta \vec V\cdot \vec n}{R_I}\frac{(1+e)\epsilon g(s)}{1+\epsilon g^2(s)}.
		\end{align}
		According to the Kramers-Moyal expansion, we obtain the differential form of the master equation as
		\begin{align}
			&\frac{\partial P(\omega,t)}{\partial t}= \gamma\left[\frac{\partial }{\partial \omega}\omega + \frac{T}{I}\frac{\partial^2 }{\partial \omega^2}\right]P(\omega,t)\notag\\
				&+\sum_{n=1}^\infty \frac{(-1)^n}{n!}\frac{\partial^n }{\partial \omega^n}K_n(\omega)P(\omega,t),
		\end{align}
		where we have introduced the Kramers-Moyal coefficients
		\begin{align}
			&K_n(\omega)		= \int ds \int d\vec v (\Delta\omega)^n \rho h f(\vec v)\Theta(\Delta \vec V\cdot \vec n)|\Delta \vec V\cdot \vec n| \notag\\
							&= \rho h\!\!\int ds\!  \left[\frac{\epsilon (1+e)g(s)}{R_I(1+\epsilon g^2(s))}\right]^n
								\!\!\int \!\!d\vec v f(\vec v)\Theta(\Delta \vec V\cdot \vec n)(\Delta \vec V\cdot \vec n)^{n+1}.
		\end{align}

	\subsection{Small noise expansion}
		We consider the following four assumptions: 
		(i) {\it $\epsilon$ is a small positive parameter,}
		(ii) {\it $\gamma$ is a small positive number independent of $\epsilon$,}  
		(iii) {\it $T$ is scaled as $T=\epsilon^2\mc{T}$, where $\mc{T}$ is independent of $\epsilon$, } and
		(iv) {\it $f(\vec v)$ is isotropic as $f(\vec v)=\phi(|\vec v|)$.}
		Introducing a scaled variable $\hat \omega=\epsilon\hat \Omega$, we obtain the scaled master equation as 
		\begin{align}
			\frac{\partial \mc{P}(\Omega,t)}{\partial t}&= \gamma \left[\frac{\partial }{\partial \Omega}\Omega+\frac{\mc{T}}{I}\frac{\partial^2 }{\partial \Omega^2}\right]\mc{P}(\Omega,t) \notag\\
				&+\sum_{n=1}^\infty \frac{(-1)^n}{n!}\frac{\partial^n }{\partial \Omega^n}\mathcal{K}_n(\Omega)\mc{P}(\Omega,t), \label{eq:Suppl:scaled_ME}
		\end{align}
		with the scaled Kramers-Moyal coefficients 
		\begin{align}
			&\mathcal{K}_n(\Omega) = \rho h\int ds  \frac{(1+e)^ng^n(s)}{R_I^n(1+\epsilon g^2(s))^n}\times \notag\\
			\int d\vec v &\phi(|\vec v|)\Theta((\epsilon \vec {\mathcal{V}}(s)-\vec v)\cdot \vec n)[(\epsilon\vec {\mathcal{V}}(s)-\vec v)\cdot \vec n]^{n+1}, 
		\end{align}
		where $\mathcal{\mathcal{V}}=\Omega\vec e_z \times \vec r(s)$. 
		In the limit $\epsilon\rightarrow+0$, Eq.~{(\ref{eq:Suppl:scaled_ME})} is reduced to 
		\begin{align}
			\frac{\partial \mc{P}(\Omega,t)}{\partial t}&= \gamma \left[\frac{\partial }{\partial \Omega} \Omega+\frac{\mc{T}}{I}\frac{\partial^2 }{\partial \Omega^2}\right]\mc{P}(\Omega,t) \notag\\
				&+\sum_{n=1}^\infty \frac{(-1)^n}{n!}\frac{\partial^n }{\partial \Omega^n}\mathcal{K}_n\mc{P}(\Omega,t), 
		\end{align}
		\begin{equation}
			\mathcal{K}_n = \frac{\rho h(1+e)^n}{R_I^n}\int ds g^n(s)
								\int d\vec v \phi(|\vec v|)\Theta(-\vec v\cdot \vec n)(-\vec v\cdot \vec n)^{n+1}. 
		\end{equation}
		Here we can calculate the integral with the aid of
		\begin{align}
			&\int d\vec v \phi(|\vec v|)\Theta(-\vec v\cdot \vec n)(-\vec v\cdot \vec n)^{n+1}\notag\\
			=& \int_0^\infty dvd\theta d\psi v^2\sin{\psi}   \phi(v)\Theta(-v\cos{\psi})(-v\cos{\psi})^{n+1}\notag\\
			=& 2\pi\int_0^\infty dvv^{n+3}\phi(v)\int_{\pi/2}^{\pi} (-\cos{\psi})^{n+1}\sin{\psi}d\psi\notag\\
			=& \frac{2\pi}{n+2}\int_0^\infty dvv^{n+3}\phi(v),
		\end{align}
		and
		\begin{align}
			&\int ds g^n(s) 
			= \frac{2}{R_I^{n}}\int_{-l/2}^{l/2} ds' s'^n + \frac{2}{R_I^{n}}\int_{-w/2}^{w/2} ds' s'^n \notag\\
			&= 	\begin{cases}
					\frac{4}{R_I^n(n+1)}\left[\left(\frac{l}{2}\right)^{n+1} \!+\! \left(\frac{w}{2}\right)^{n+1}\right] & ({\rm for\>\>even\>\>}n) \cr
					0 & ({\rm for\>\>odd\>\>}n)
				\end{cases}.
		\end{align}
		where we have used the isotropic distribution $\phi(\vec v)=\phi(v)$, and 
		decomposed the position vector $\vec r$ as $\vec r=x\vec e+s'\vec t+z\vec e_z$ with $x=\pm l/2$ or $x=\pm w/2$. 
		We thus have the relation $g=s'/R_I$. 
		Then, the cumulant $\mathcal{K}_n$ is simplified as 
		\begin{equation}
			\mc{K}_n \!=\!	\begin{cases}
							\frac{4\pi \rho h(1+e)^n (l^{n+1} + w^{n+1})}{2^n R_I^{2n}(n+1)(n+2)}\!\!\int_0^\infty \!\!dvv^{n+3}\phi(v) & \!\!({\rm for\>\>even\>\>}n) \cr
							0 & \!\!({\rm for\>\>odd\>\>}n)
						\end{cases},\label{BL_cum}
		\end{equation}
		which implies that the cumulant function is given as 
		\begin{align}
			&\Phi(s) +\mc{T}\!s^2 =   \sum_{n=1}^\infty \frac{(is)^n}{n!}K_n\notag\\
					&= -\!\frac{16\pi \rho hR_I^4}{ls^2(1\!+\!e)^2}\int_0^\infty \!\!\!\!dv v\phi(v) \sum_{n=2}^\infty \frac{(-1)^n}{(2n)!}\left[\frac{s(1+e)lv}{2R_I^2}\right]^{2n}\notag\\
					   &-\frac{16\pi \rho hR_I^4}{ws^2(1\!+\!e)^2}\int_0^\infty\!\!\!\! dv v\phi(v) \sum_{n=2}^\infty \frac{(-1)^n}{(2n)!}\left[\frac{s(1+e)wv}{2R_I^2}\right]^{2n}\notag\\
					&= -\!\frac{8\pi \rho hR_I^4}{ls^2(1\!+\!e)^2}\int_{-\infty}^\infty \!\!\!\!\!\!\!dv |v|\phi(v) \!\!\left[e^{\frac{is(1+e)lv}{2R^2_I}}\!\!\!\!\!-1+\frac{s^2(1\!+\!e)^2l^2v^2}{8R^4_I}\right]\notag\\
					   &\!-\!\frac{8\pi \rho hR_I^4}{ws^2(1\!+\!e)^2}\int_{-\infty}^\infty \!\!\!\!\!\!\!dv |v|\phi(v) \!\!\left[e^{\frac{is(1+e)wv}{2R^2_I}}\!\!\!\!\!-1+\frac{s^2(1\!+\!e)^2w^2v^2}{8R^4_I}\right].\label{eq:Suppl:cumulant_general}
		\end{align} 
		
	\subsection{In the case with the exponential velocity distribution}
		Let us consider the case of $\mc{T}=l=0$, and assume that the velocity distribution of the granular gas is given by the exponential form as
		\begin{equation}
			f(\vec v) = \frac{1}{8\pi v_0^3}e^{-|v|/v_0},
		\end{equation}
		with the characteristic velocity $v_0$. 
		The non-Gaussian Langevin equation for the scaled angular velocity $\hat \Omega\equiv\omega/\epsilon$ is 
		\begin{equation}
			\frac{d\hat{\Omega}}{dt} = -\gamma \hat{\Omega} + \hat {\eta}_g.
		\end{equation}
		Here, the cumulants function is given by 
		\begin{equation}
			\Phi(s) = -\frac{\rho hwv_0\Omega_g^2s^2 (5+3\Omega_g^2s^2)}{2(1+\Omega^2_gs^2)^2},
		\end{equation}
		where we have introduced $\Omega_g\equiv wv_0(1+e)/2R_I^2$.  
		Then, we obtain the velocity distribution (11) in the main text for the scaled angular velocity $\tl{\Omega}\equiv \Omega/\Omega_g$ as 
		\begin{equation}
			\mc{P}_{\rm SS}(\tl{\Omega}) = \int_{-\infty}^\infty \frac{ds}{2\pi} \frac{1}{(1+s^2)^{3v_0/2\tl{v}}}\exp{\left[-is\tl{\Omega} - \frac{v_0s^2}{\tl{v}(1+s^2)}\right]},
		\end{equation}
		where $\mc{P}_{\rm SS}(\tl{\Omega})\equiv \mc{P}_{\rm SS}(\Omega)\Omega_g$, and $\tl{v}\equiv 2\gamma/\rho hw$
		\begin{figure*}
			\includegraphics[width=175mm]{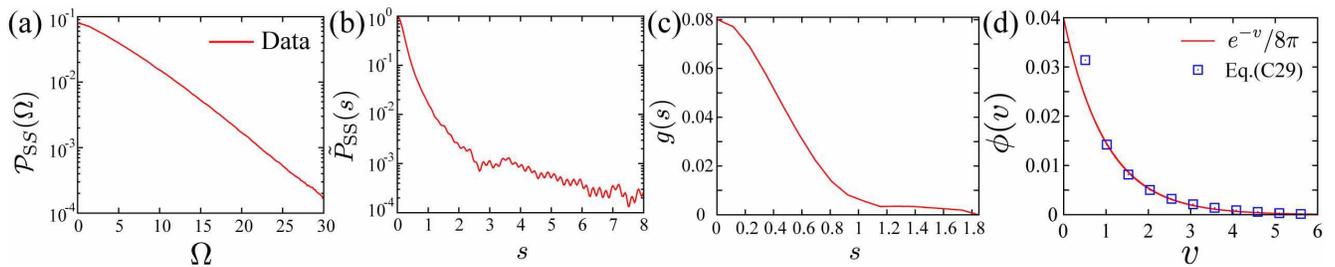}
			\caption{
						Numerical data of the Monte Carlo simulation of Eq.~{(\ref{eq:Suppl:BLeq})}. 
						(a)~The steady distribution function of rotor's angular velocity. 
						(b)~The numerical Fourier transform of $P_{\rm SS}(\Omega)$. 
						(c)~The numerical data of $g(s)$. If we ignore the numerical fluctuation, $g(s)$ tends to converge to zero in the limit of $s\rightarrow+\infty$. 
						(d)~The estimated data of the granular velocity distribution using Eq.~{(\ref{spp:eq:estdist_num})}. 
						    Because of the singularity at $v=0$ in Eq.~{(\ref{spp:eq:estdist_num})}, the accuracy of the data near $v=0.5$ is not good.
					}
			\label{fig:Suppl:dist}
		\end{figure*}
	\subsection{Inverse estimation formula for the spherical distribution}
		We derive the inverse formula of the granular velocity distribution for the case of $\mc{T}=l=0$ and an arbitrary $\phi(v)$. 
		The non-Gaussian Langevin equation is given by 
		\begin{equation}
			\frac{d\hat{\Omega}}{dt} = -\gamma\hat \Omega +  \hat \eta_g, 
		\end{equation}
		where the cumulant function of $\hat{\eta}_{g}$ is given by
		\begin{equation}
			\Phi(s) = -\frac{2\pi \rho hw}{s^2F_g^2}\int_{-\infty}^\infty dv |v|\phi(v) \left[e^{iF_gsv}-1+\frac{F_g^2s^2v^2}{2}\right],
		\end{equation}
		with the typical collisional impact $F_g\equiv w(1+e)/2R_I^2$. 
		From Eq.~{(\ref{spp:g_Fdist})}, we obtain  the following relation between the granular velocity distribution and the Fourier representation of rotor's angular velocity distribution as
		\begin{align}
			&-\frac{2\pi \rho hw}{s^2F_g^2}\int_{-\infty}^\infty dv |v|\phi(v) \left[e^{iF_gsv}-1+\frac{F_g^2s^2v^2}{2}\right] \notag\\
			&=  \gamma s\frac{d}{ds}\log{\tl{P}_{SS}(s)}.\label{eq:Suppl:est1}
		\end{align}
		This formula can be transformed into the following form:
		\begin{equation}
			\phi(v) \!=\! \frac{1}{\pi |v|} \!\int_{0}^\infty \!\!ds \!\left[a\!-\!\frac{bs^2}{2}\!-\!cs^3\frac{d}{ds}\log{\tl{P}_{\rm SS}(s/F_g)}\right]\cos{(sv)},\label{eq:Suppl:est2}
		\end{equation}
		where we introduced $a\equiv\int_{-\infty}^{\infty}dv|v|\phi(v)$, $b\equiv\int_{-\infty}^{\infty}dv|v|^3\phi(v)$, and $c\equiv \gamma/2\pi\rho hw$. 

		Let us explain how to determine the coefficients $a$ and $b$. 
		According to the Riemann-Lebesgue lemma~{\cite{RLlemma}}, the following relation holds if $|v|\phi(v)$ is an $L^1$-function: 
		\begin{equation}
			\lim_{s\rightarrow+\infty}\int_{-\infty}^{\infty} ds |v|\phi(v)e^{isv} = 0,
		\end{equation}
		or equivalently, 
		\begin{equation}
			\lim_{s\rightarrow\infty}\left[a-\frac{bs^2}{2}-cs^3\frac{d}{ds}\log \tl{P}_{\rm SS}(s/F_g)\right] = 0. \label{eq:Suppl:RLlemma}
		\end{equation}
		Equation~{(\ref{eq:Suppl:RLlemma})} is practically useful to determine the coefficients $a$ and $b$ from the experimental data of $\tl{P}_{\rm SS}(s)$.
		
	\subsection{The numerical technique for the inverse estimation formula}
		Here we explain our numerical procedure for the inverse formula~{(\ref{eq:Suppl:est2})}. 
		We obtain the steady distribution function of rotor's angular velocity using the Monte Carlo simulation of Eq.~{(\ref{eq:Suppl:BLeq})} on the following setup: $\phi(v)=e^{-|v|}/8\pi$, $l=T=0$, $w=\sqrt{12}$, $M=\rho=h=e=I=1$, $m=0.01$, and $\gamma=2$. 
		The numerical data is plotted in Fig.~{\ref{fig:Suppl:dist}(a)}. 
		
		To obtain the Fourier transformation $\tl{P}_{\rm SS}(s)$, we have used the numerical distribution $\mc{P}_{SS}(\Omega)$ for $0\leq \Omega\leq 30$. 
		$\tl{P}_{\rm SS}(s)$ is numerically plotted in Fig.~{\ref{fig:Suppl:dist}(b)}. 
		We numerically estimate the coefficients $a$ and $b$ as $a=0.080121$ and $b=0.464$, respectively, 
		and obtain the following function as shown in Fig.~{\ref{fig:Suppl:dist}(c)}: 
		\begin{equation}
			g(s)	=a - \frac{bs^2}{2} - cs^3\frac{d}{ds}\log{\tl{P}_{\rm SS}(s/F_g)}. 
		\end{equation}
		Figure~{\ref{fig:Suppl:dist}(c)} implies the asymptotic form $g(s) \simeq 0$ for $s\rightarrow\infty$. 
		On the basis of the numerical data of $g(s)$, 
		we estimate the granular velocity distribution function $\phi(v)$ as
		\begin{equation}
			\phi(v) = \frac{1}{\pi |v|}\int_0^\infty dsg(s)\cos{sv}.\label{spp:eq:estdist_num}
		\end{equation}
		We plot the granular velocity distribution estimated from Eq.~{(\ref{spp:eq:estdist_num})} in Fig.~{\ref{fig:Suppl:dist}(d)}. 
		We note that Eq.~{(\ref{spp:eq:estdist_num})} has a singularity at $v=0$, which explains that the numerical accuracy is not good around $v=0$.

\end{document}